\documentclass[prd,showpcs,amsmath,amssymb,nofootinbib,longbibliography,twocolumn,showpacs,notitlepage,superscriptaddress
]{revtex4-1}
\usepackage{multirow}
\usepackage{epsfig}
\usepackage{amsmath}
\usepackage{bm}
\usepackage{times}
\usepackage{graphicx}
\usepackage{color}
\usepackage{slashed}
\usepackage{graphicx}
\usepackage{amsmath}
\usepackage[latin1]{inputenc}
\usepackage{hyperref}
\usepackage{soul}
\usepackage{multirow}
\usepackage{epsfig}
\usepackage{amsmath}
\usepackage{bm}
\usepackage{times}
\usepackage{graphicx}
\usepackage{color}
\usepackage{slashed}
\usepackage{graphicx}
\usepackage{amsmath}
\usepackage{tikz}
\usetikzlibrary{positioning,shapes}
\usepackage{relsize} 
\usepackage[latin1]{inputenc} 
\usepackage{tikz}
\usepackage{gensymb}
\usetikzlibrary{trees}
\usetikzlibrary{decorations.pathmorphing}
\usetikzlibrary{decorations.markings}
\usetikzlibrary{positioning,arrows}
\usetikzlibrary{decorations.pathmorphing}
\usetikzlibrary{decorations.markings}
\usetikzlibrary{decorations.pathreplacing,calc}
\usetikzlibrary{decorations.pathmorphing,decorations.markings,trees,positioning,arrows}   
\newif\ifmirrorsemicircle
\usetikzlibrary{decorations.pathreplacing,decorations.markings,arrows}

\pgfarrowsdeclarecombine[-5pt]{circled}{circled}{latex}{latex}{o}{o}
\pgfarrowsdeclaredouble{doubled}{doubled}{stealth}{stealth}

\def\bea{\begin{eqnarray}}
\def\eea{\end{eqnarray}}
\def\bean{\begin{eqnarray*}}
\def\eean{\end{eqnarray*}} 

\def\beaal{\begin{align}}
\def\eeaal{\end{align}}

\begin{document}  
 
\title{Dark Matter Capture by Atomic Nuclei}

\author{Bartosz~Fornal}
\affiliation{Department of Physics and Astronomy, University of Utah, Salt Lake City, UT 84112, USA\vspace{1mm}}
\author{Benjam\'{i}n~Grinstein}
\affiliation{\vspace{1mm}Department of Physics, University of California, San Diego, 9500 Gilman Drive, La Jolla, CA 92093, USA\vspace{0mm}} 
\author{Yue~Zhao\vspace{1mm}}
\affiliation{Department of Physics and Astronomy, University of Utah, Salt Lake City, UT 84112, USA\vspace{1mm}}
\date{\today}

\begin{abstract}
\vspace{0mm}We propose a new strategy to search for a particular type of dark matter via nuclear capture. If the dark matter particle carries baryon number, as motivated by a class of theoretical explanations of the matter-antimatter asymmetry of the universe, it can mix with the neutron and  be captured by an atomic nucleus. The resulting state de-excites by emitting a single photon or a cascade of photons with a total energy of up to several MeV. The exact value of this energy depends on the dark  matter mass.  We investigate the prospects for detecting dark matter capture signals in current and future neutrino and  dark matter direct detection experiments. \\ \vspace{0mm}
\end{abstract}

\maketitle

\section{Introduction}
Two of the greatest mysteries of modern particle physics are the nature of dark matter and the origin of the matter-antimatter asymmetry of the universe. Among the plethora of existing theoretical models, minimal extensions of the Standard Model addressing both of those questions simultaneously are of particular interest.

One of the simplest interactions of a fermionic dark sector particle $\chi$ (a gauge singlet) with the baryonic sector can be effectively described by a dimension six operator
\bea\label{op}
\mathcal{O} \sim \frac{q q q \bar\chi}{\Lambda^2} \ ,
\eea
where $\Lambda$ denotes the scale of new physics mediating this interaction, e.g., the mass of a heavy color triplet scalar mediator.\break A possible realization of this  operator  is $\epsilon^{ijk} \overline{u^c_{i}}_{\!R}d_{jR}d_{kR}\bar{\chi}/\Lambda^2$. The dark sector particle $\chi$ carries baryon number $B = 1$.

The existence of such an interaction has  been proposed as a possible solution to the neutron lifetime anomaly  \cite{Fornal:2018eol}. The anomaly arises from the fact that experiments measuring the free neutron lifetime by determining the change in the number of neutrons in time yield a lifetime shorter than experiments sensitive only to protons in the final state. The two results can be reconciled if the neutron has a new \emph{dark decay} channel with a $\sim 1\%$ branching fraction. 

It has been argued \cite{McKeen:2018xwc,Baym:2018ljz,Motta:2018rxp} that such a dark decay channel, in the absence of additional interactions involving the dark particle, would allow neutron stars to reach masses only up to $\sim 0.8 \, {\rm M_\odot}$, substantially below  the observed cases with masses $2 \, M_\odot$. This problem, however, is solved either by introducing self-interactions  in the dark sector \cite{Cline:2018ami,Karananas:2018goc} or an additional repulsive interaction between the dark matter and the neutron \cite{Grinstein:2018ptl}. Interestingly, it has also been shown that models with neutron dark decay offer a successful framework for explaining the matter-antimatter asymmetry of the universe \cite{Bringmann:2018sbs}.

One of the possible neutron dark decay channels involves a dark sector particle $\chi$ and a photon in the final state, i.e.,\break $n\to \chi\gamma$. This  channel has been searched for directly in experiment and a $1\%$ branching fraction  is in tension with data \cite{Tang:2018eln}. Part of the available parameter space is further constrained by the results of the Borexino experiment \cite{Agostini:2015oze} translated into a limit on the hydrogen lifetime \cite{11}.

Two particularly appealing scenarios are  realized when the dark particle $\chi$ from the neutron dark  decay $n\to\chi\gamma$ is the dark matter or the antiparticle of dark matter. In the latter case, which  has been considered in Refs.\,\cite{Jin:2018moh,Keung:2019wpw}, $\bar\chi$ is then the dark matter and carries baryon number $B=-1$. It can, therefore, annihilate with nucleons, leading to spectacular signatures in various experiments.

In this letter we focus on the first possibility, i.e., when the dark particle $\chi$ is the dark matter, and we propose a complementary method to search for $\chi$. Since in this case $\chi$ carries baryon number $B=1$, as the Earth moves through the dark matter halo in our galaxy, $\chi$  can be captured by atomic nuclei through its mixing with the neutron. 
 We explore the prospects of using large volume neutrino experiments and dark matter detectors to look for such processes.

\section{Preliminaries}

The operator $\epsilon^{ijk} \overline{u^c_{i}}_{\!R}d_{jR}d_{kR}\bar{\chi}/\Lambda^2$ induces mixing between the dark matter  particle $\chi$ and the neutron. 
At the hadron level, such a theory, including also the neutron magnetic moment interaction,  is described by the effective Lagrangian
 \bea\label{lageff113}
\mathcal{L}_{\rm eff} &=&\bar{n}\,\Big(i\slashed\partial-m_n +\frac{g_ne}{8 m_n}\sigma^{\,\mu\nu}F_{\mu\nu}\Big) \,n\nonumber\\
&+&  \bar{\chi}\,\big(i\slashed\partial-m_\chi\big) \,\chi + \varepsilon \left(\bar{n}\,\chi + \bar{\chi}\,n\right) \ ,
\eea
where the model-dependent mixing parameter $\varepsilon \sim b/\Lambda^2$ (with $b \!=\! 0.0144(3)(21) \ {\rm GeV^3}$ \cite{Aoki:2017puj}) and $g_n$ is the neutron $g$-factor. 
One of the possible particle physics realizations  involves a heavy color triplet scalar mediating the mixing between the dark matter and the neutron \cite{Fornal:2018eol}.
The resulting dark matter mass eigenstate contains a small  admixture of the neutron,
\bea\label{3}
|\chi\rangle' = |\chi\rangle +\frac{\varepsilon}{m_n-m_\chi} \,|n\rangle 
\eea
and vice versa. 
In this scenario, if energetically allowed, the neutron decays to $\chi$ and a photon at a rate 
\bea\label{rate}
\Delta\Gamma_{n\rightarrow \chi\gamma} = \frac{g_n^2e^2}{128\pi}\bigg(1-\frac{m_\chi^2}{m_n^2}\bigg)^3  \frac{m_n\,\varepsilon^2}{(m_n-m_\chi)^2}  \ .
\eea
A branching fraction for this dark decay channel at the level of $1\%$ would explain the neutron lifetime anomaly.
The allowed dark matter mass  range  is
\bea\label{one2}
937.993 \ {\rm MeV}< m_\chi < 938.783 \ {\rm MeV} \ ,
\eea
where the lower bound assures that none of the stable nuclei undergo dark decays, whereas the upper bound is necessary for the stability of $\chi$  (with respect to $\beta$ decay).

\begin{figure}[t!]
\includegraphics[trim={2.5cm 0.5cm 0 0},clip,width=10cm]{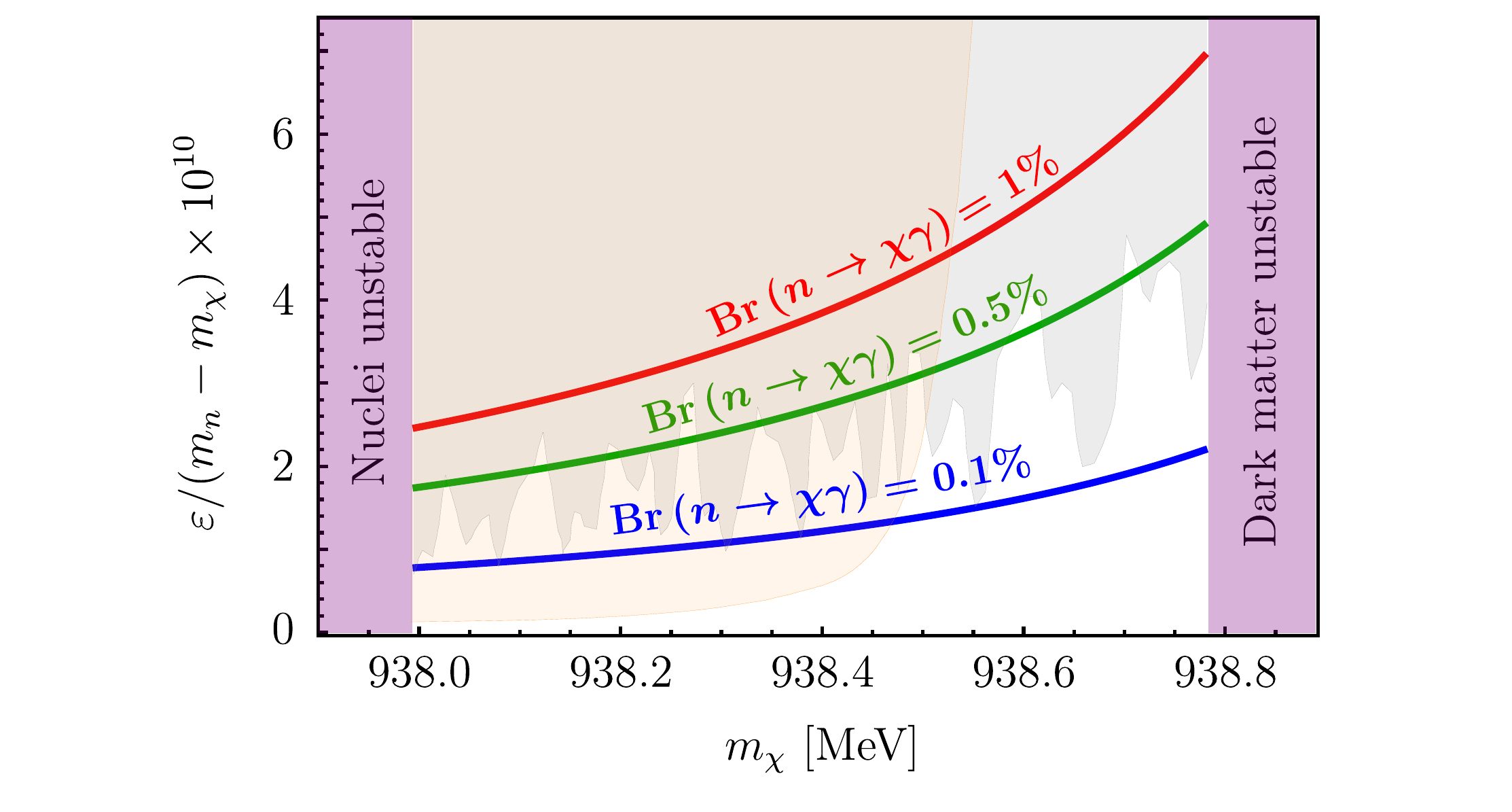}
\caption{The values of $\varepsilon/(m_n-m_\chi)$ in units of $10^{-10}$ (as a function of the dark matter mass $m_\chi$) that yield the
neutron dark decay branching fractions: $1\%$ (red curve), $0.5\%$ (green) and $0.1\%$ (blue).\break  For dark matter masses $m_\chi < 937.993 \ {\rm MeV}$ stable nuclei would undergo dark decays, whereas for  $m_\chi > 938.783 \ {\rm MeV}$ the dark matter $\chi$ would be unstable with respect to $\beta$ decays. The boundaries of the orange and gray-shaded regions are the upper limits on $\varepsilon/(m_n-m_\chi)$ derived from the Borexino data in Ref.\,\cite{11} and from the direct search for $n\to \chi\gamma$ \cite{Tang:2018eln}, respectively, both at the $90\%$ confidence level.\\}\label{fig1}
\end{figure}

Figure \ref{fig1} shows the values of the dimensionless parameter\break  $\varepsilon/(m_n\!-\!m_\chi)$ that yield the neutron dark decay $n\to\chi\gamma$ branching fractions: $1\%$ (red curve), $0.5\%$ (green) and $0.1\%$ (blue),
for dark matter masses in the  range specified in Eq.\,(\ref{one2}). The boundaries of the orange and gray-shaded regions correspond to the $90\%$ confidence level upper limits on\break $\varepsilon/(m_n\!-\!m_\chi)$ based on the analysis of the low-energy photon spectrum of the Borexino data conducted in Ref.\,\cite{11} and the direct search for $n\to \chi\gamma$ \cite{Tang:2018eln}, respectively.

\section{Dark matter capture}\label{sec3}

Thus far, it has not been appreciated that in models with a mixing between the dark matter and the neutron, it is possible for the dark matter particle to be captured by atomic nuclei.
This process can be interpreted as a nuclear capture of an off-shell neutron with a mass and kinetic energy equal to those of the dark matter particle. 

The capture of $\chi$  on a  nucleus  $(A,Z)$ leads to
\bea
\chi + (A,Z) \to   (A+1, Z)^* \to (A+1,Z)  + \gamma_c \ , \ \ \ 
\eea
where $\gamma_c$ denotes  a single photon or a cascade of photons  from the de-excitation of $(A+1, Z)^*$ to the ground state. 
For nonrelativistic $\chi$ $(v_\chi \ll c)$, the total energy of the  cascade is
\bea\label{energyA}
E_c = S(n) - (m_n - m_\chi) \ ,
\eea
where $S(n)$ is the neutron separation energy in $(A+1,Z)$, and we neglected  the kinetic energy of $\chi$.
Given Eq.\,(\ref{3}), the cross section for dark matter capture can be written as 
\bea\label{epsilon2}
\sigma_{\chi} = \sigma_{n^*} \ \frac{\varepsilon^2}{(m_n-m_\chi)^2}\ ,
\eea
where $\sigma_{n^*}$ is the cross section for the capture of an off-shell neutron, or, approximately, a particle with identical properties as the neutron, but with mass $m_\chi$ and kinetic energy $E_k$ equal to the kinetic energy of $\chi$.

In the energy region of the capture state  far away from resonances, the only contribution to the cross section $\sigma_{n^*}$ comes from the nonresonant (NR) channel capture. This contribution was calculated for standard neutron capture in Ref.\,\cite{LANE1960563} (see also \cite{MUGHABGHAB197993}), and in our case  it can be  written as
\bea\label{99}
\sigma_{n^*}^{\rm NR}  &=& \frac{0.062}{R\sqrt{E_k}}\frac{Z^2}{A^2} \sum_f \mu\frac{2J_f+1}{6(2I+1)}S_{dp} \, W(y_f) \ ,
\eea
where
\bea\label{WW}
W(x) = \frac{(x+3)^2}{(x+1)^2}\,x^2\left[1+\frac{(R-a_s)}{R}\frac{x(x+2)}{x+3}\right]^2. \ \ \ \ \ \ \ 
\eea
The sum is over all the final states resulting from the emission of the first  photon in each  cascade.
In the above equations $I$, $J_f$, $S_{dp}$, $a_s$ and $R$ are: the  target spin, the final state spin, the spectroscopic factor, the coherent scattering length (roughly equal to the potential scattering radius) and the nuclear radius, respectively. The variable $\mu$ takes the following values:  if $I=0$ then $\mu=1$; if $I\ne0$ then $\mu=1$ for $J_f = I\pm3/2$ and  $\mu=2$ for $J_f = I\pm1/2$. The parameter $y_f$
is given by
\bea
y_f = \frac{\sqrt{2m_\chi E_f}}{\hbar} \, R \ ,
\eea
where $E_f$ is the energy of the first emitted photon.

The first term inside the brackets  in Eq.\,(\ref{WW}) corresponds to hard sphere capture, whereas the second term represents the contribution  from distant resonances.  
Since $m_\chi\approx m_n$, the only  difference between our case and neutron capture is in the values of $E_f$, which are related via
\bea
E_f^{(\chi \ {\rm capture})} = E_f^{(n \ {\rm capture})} - (m_n - m_\chi) \ .
\eea
For $E_f^{(n \ {\rm capture})}$ on the order of several $\rm MeV$, the cross section  $\sigma_{n^*}^{\rm NR}$  is not very different from the nonresonant channel cross section of standard neutron capture $\sigma_{n}^{\rm NR}$.

If the capture state energy is within the resonance region, there are additional contributions to the  cross section that depend on the resonance structure. In the case of dark matter capture, the energy of the capture state is below the threshold for neutron emission, i.e., it is in the region of bound states. The exact contribution of the resonant capture to the cross section  depends on how well the state resulting from the capture of a dark matter particle with a given kinetic energy matches the energy of the nuclear state. Consequently, the dark matter capture cross section will depend on the velocity distribution of the particles being captured, on the density of states in the capture energy region, and on the widths of the nuclear states.

For bound states the only available de-excitation channels are through photon emission, which implies that the states have widths $\sim 1 \ {\rm eV}$ or smaller. In later discussion, we mainly focus on argon and xenon as our target elements. The excitation energies range from 5 to 8 MeV, and the densities of nuclear states are on the order  of $40/{\rm MeV}$ for $^{41}{\rm Ar}$ (the spacing between the states is $\sim 25 \ {\rm keV}$) and  $20\,000/{\rm MeV}$ for $^{132}{\rm Xe}$ (the spacing  is $\sim 50 \ {\rm eV}$).

Regarding the kinetic energy of the dark matter particle $\chi$ being captured, within the Standard Halo Model \cite{Drukier:1986tm}  the dark matter velocity distribution with respect to the center of the Milky Way can be approximated by the Maxwell-Boltzmann distribution
\bea\label{distr}
f(\vec{v}) =
\begin{cases}
 \frac{1}{N_{\rm esc}} \left(\frac{3}{2\pi \sigma_v^2}\right)^{\frac32} \exp\left({-\frac{3\vec{v}^2}{2\sigma_v^2}}\right) \ \ \ {\rm for} \ \ |\vec{v}|<v_{\rm esc} \ , \ \ \ \  \vspace{2mm} \\
\ \ 0 \ \ \ \ {\rm for} \ \ \ |\vec{v}|>v_{\rm esc} \ ,
\end{cases}
\eea
with $\sigma_v$ being the root mean square velocity dispersion and $N_{\rm esc} = {\rm Erf}(v_{\rm esc}/v_0)-2v_{\rm esc}\exp(-v_{\rm esc}^2/v_0^2)/(\sqrt{\pi}\,v_0)$, where $v_0 = \sqrt{2/3}\,\sigma_v \approx 235 \ {\rm km/s}$ is the most probable  velocity and $v_{\rm esc} \approx 500-600 \ {\rm km/s}$ is the escape velocity.

Since the Solar System is moving around the center of the Milky Way with a speed   $v_{\oplus} \approx220 \ {\rm km/s}$ \cite{Kerr:1986hz}, the expected dark matter velocity distribution  measured on Earth is given by $f(\vec{v}-\vec{v}_\oplus)$. The most probable dark matter velocity shifts to $\sim 320 \ {\rm km/s}$. 
Translating this into the dark matter kinetic energy distribution, the most probable kinetic energy of $\chi$ (measured on Earth) is $\sim 530 \ {\rm eV}$ and the half-width of the distribution is $\sim 1100 \ {\rm eV}$.

An additional contribution to the kinetic energy distribution of $\chi$ comes from the Earth's motion within the Solar System. 
The Earth's velocity around the Sun is $\sim 30 \ {\rm km/s}$ and its orbit is inclined at an angle $\sim 60\textdegree$ relative to the plane of the Milky Way. Therefore, the kinetic energy distribution of $\chi$ in experiments on Earth will shift  by $\sim \pm \,50 \ {\rm eV}$ over the course of the year. For a detailed discussion of the resulting annual modulation of dark matter signals, see Ref.\,\cite{RevModPhys.85.1561}.

Given the large width of the dark matter kinetic energy distribution and the narrowness of the nuclear bound states, the resonant $\chi$ capture does not dramatically increase  the total capture cross section above its nonresonant value. 
To make a rough estimate of this enhancement, consider the case of xenon. Assuming a spacing between the states of  $\sim 50 \ {\rm eV}$, their width $\sim 1 \ {\rm eV}$ and an average enhancement within a resonance over  the nonresonant contribution  of $\sim 1000$ (which corresponds to the enhancement theoretically predicted for neutron capture by $^{131}{\rm Xe}$ at $E_k \sim 500 \ {\rm eV}$), the resonant contribution to the dark matter capture cross section is only $\sim 10$ times larger than the nonresonant contribution.

A nontrivial resonant structure would also lead to a varying cross section for dark matter capture throughout the year due to the shift in the kinetic energy distribution of $\chi$, thus an annual modulation of the signal would be expected.
We\break leave a detailed analysis of these effects for future work; the results presented in this study provide a conservative estimate for the dark matter capture signal.

The rate of dark matter capture is calculated as
\bea\label{rrate}
R =  n\int d^3\vec{v} \, f(\vec{v}-\vec{v}_{\oplus}) \,\Phi_\chi(v)  \, \sigma_\chi(v) \ ,
\eea
where $\Phi_\chi$ is the dark matter flux  and $n$ is the number density of target nuclei. Given the dark matter  density in the vicinity of the Solar System of $\sim 0.3 \ {\rm GeV/cm^3}$, for a dark matter with mass in the range given in Eq.\,(\ref{one2}),  the  flux (governed by the velocity distribution of $\chi$) has a maximal value 
$
\Phi_\chi(v_0) \approx\, 10^7 \ \, {{\rm cm}^{-2} \,{\rm s}^{-1}} 
$.
The number density $n$ for a detector filled with a substance of mass $M$ and mole density $\rho_{\rm mol}$ is
$
n = {N_0 M}/{\rho_{\rm mol}} \,,
$
where $N_0$ is the Avogadro number.\\

\section{Signals in  experiments}

Signatures of dark matter capture can be searched for in large volume neutrino experiments and dark matter direct detection experiments which are capable of recording $\rm MeV$ photon energy depositions.
We first investigate the possible constraints on $\chi$ capture by hydrogen arising from the Borexino experiment \cite{Alimonti:2008gc}.  We then discuss the unique signatures of $\chi$ capture by argon at the future Deep Underground Neutrino Experiment (DUNE) \cite{Abi:2020wmh} and investigate the prospects of discovering $\chi$ capture by xenon at PandaX \cite{Cao:2014jsa}, XENON1T \cite{Aprile:2015uzo} and LUX \cite{Akerib:2012ys}. Finally, we comment on the sensitivity of other existing and upcoming neutrino and dark matter experiments, and propose detector materials that would increase the dark matter capture signal.

\subsection*{Borexino}

Borexino recorded the photon  spectrum of low-energy neutrinos interacting with electrons \cite{Agostini:2017ixy}. The detector contained  $\sim 70$ tons of pseudocumene (made up of hydrogen, carbon and oxygen) and the exposure time was $\sim 1300$ days. 

The cross sections for radiative neutron capture by those constituents at energies $E_k\sim 530 \ {\rm eV}$ are \cite{etde_519804}:  $\sigma_n(^1{\rm H}) \approx 2 \ {\rm mb}$, $\sigma_n(^{12}{\rm C}) \approx 20 \ {\rm \mu b}$ \,and\, $\sigma_n(^{16}{\rm O}) \approx 4 \ {\rm \mu b}$. Since the rate of dark matter capture by hydrogen is several orders of magnitude larger  than on carbon and oxygen,  in the following analysis of Borexino data we focus solely on hydrogen (we estimate its mass in the detector to be $\sim 7$ tons).

Neutron capture by hydrogen and subsequent de-excitation of deuterium lead to the emission of a single photon with energy $E_\gamma^{(n)}(^1{\rm H}) = 2.225 \ {\rm MeV}$ \cite{MONAHAN1961400}. This implies that the signature of dark matter capture by hydrogen,
\bea
\chi + {^{1}{\rm H}} \to \,{^{2}{\rm H}}^* \to \,{^{2}{\rm H}} + \gamma \ ,
\eea
consists of a single monochromatic photon with energy
\bea
E_\gamma^{(\chi)} =2.225 \ {\rm MeV} - \Delta m \ , 
\eea
where $\Delta m = m_n - m_\chi$. The photon energy  is thus uniquely determined by the dark matter mass. Equation (\ref{one2}) implies
\bea\label{rangg}
0.653 \ {\rm MeV}< E_\gamma^{(\chi)} <  1.443 \ {\rm MeV}\ .
\eea

There is no resonant behavior of the capture cross section in the case of hydrogen.
Using Eqs.\,(\ref{epsilon2}) and (\ref{99}), one arrives at
\bea\label{nstar}
\sigma_{\chi}(^1{\rm H}) = \sigma_{n}(^1{\rm H})\ \frac{W\bigg(\frac{\sqrt{2m_\chi E_\gamma^{(\chi)}}}{\hbar} \, R\bigg)}{W\bigg(\frac{\sqrt{2m_n E_\gamma^{(n)}}}{\hbar} \, R\bigg)} \ \frac{\varepsilon^2}{(m_n-m_\chi)^2}, \ \ \ \ \ \ \ 
\eea
where $W(x)$ is defined in Eq.\,(\ref{WW}).
Adopting the values of parameters for hydrogen, i.e., $R\approx 0.84 \ {\rm fm}$ \cite{Pohl:2010zza,Bezginov:2019mdi} and $a_s \approx -3.74 \ {\rm fm}$ \cite{neutrondata}, the ratio $\sigma_{n^*}^{\rm NR}/\sigma_{n}^{\rm NR}$ falls in the range $0.19 - 0.55$, with the exact value depending on the mass of $\chi$.\break  
Substituting in Eq.\,(\ref{rrate}) the velocity distribution from  Eq.\,(\ref{distr})  and the cross section from Eq.\,(\ref{nstar}), one obtains, upon setting $\sigma_{n^*}^{\rm NR}/\sigma_{n}^{\rm NR} \sim 0.4$, the expected rate of dark matter capture at Borexino of 
\bea\label{rb}
R_{B}  \ &\sim& \  3\times 10^{-5} \left[\frac{\varepsilon}{m_n-m_\chi}\times 10^{10}\right]^2  \ \ {\rm  events}/{\rm day} \ . \ \ \ \ \ \ \ 
\eea

The Borexino photon spectrum, which is in  good agreement with solar neutrino and radioactive backgrounds, can be used to derive upper bounds on $\varepsilon/(m_n-m_\chi)$. Following  the method adopted in Ref.\,\cite{Tang:2018eln}, we performed  a $Z$-test for the statistical significance of a possible dark matter capture signal in the publicly available Borexino data \cite{data}.

For photon energies $ {\sim 1 \ \rm MeV}$, the detector resolution\break is $\sim 50 \ {\rm keV}$ \cite{Agostini:2017ixy}. We combined the data bins ($\sim 2\!-\!3 \ {\rm keV}$ each) into $150 \ {\rm keV}$ segments, each with an assigned energy corresponding to the middle bin in the segment. This way, assuming the dark matter capture signal in the detector is a Gaussian of width equal to the resolution, each segment contains most of the signal corresponding to the energy of its central bin. We then calculated, for each segment (labeled by $i$) separately, the number of standard deviations between the data (with the  theoretical background  subtracted) and the predicted signal according to the $Z$-test formula
$
\Delta \sigma = {(d_i-s_i)}/{\sqrt{(\Delta d_i)^2+(\Delta s_i)^2}} \ ,
$
where $d_i$ is the number of photon counts per day with the background subtracted (provided in the publicly available Borexino data \cite{data}), $s_i$ is the signal count per day, and $\Delta d_i, \,\Delta s_i$ are the corresponding uncertainties obtained using Poisson statistics. 

Taking into account the detector's efficiency of $25\%$ \cite{Alimonti:2008gc}, we find that the parameter region $\varepsilon/(m_n-m_\chi) \lesssim 7\times10^{-9}$ is not constrained by our analysis of the Borexino data, thus the bounds   from the direct search for $n\to \chi\gamma$ \cite{Tang:2018eln} and the hydrogen lifetime \cite{11} are not improved.\vspace{1mm}

A much larger experiment which, a priori, could look for dark matter capture by hydrogen, is Super-Kamiokande  \cite{Fukuda:2002uc}. It consists of  $50$ kilotons of water and records events via Cherenkov radiation. However, the energy threshold for event selection is $3.5 \ {\rm MeV}$ \cite{Ito:2018vat}, which eliminates the potential photon signal of dark matter capture by hydrogen. Similar arguments apply to the  Sudbury Neutrino Observatory (SNO) \cite{Bellerive:2016byv}.

\subsection*{DUNE}

The DUNE detector will be based on liquid argon time\break projection chamber (LArTPC) technology, which offers unprecedented accuracy in reconstructing energies and positions of photon energy depositions. A study of the reach for MeV photons  in a LArTPC environment was conducted in the context of   the ArgoNeuT experiment and demonstrated very promising results \cite{Acciarri:2018myr}.

DUNE is planned to contain $40$ kilotons of liquid  argon. At kinetic energies $E_k \sim 530 \ {\rm eV}$  the radiative neutron capture cross section  is $\sigma_n({^{40}{\rm Ar}})\approx 5 \ {\rm mb}$ \cite{etde_519804}. The dark matter capture by {$^{40}$Ar} would  lead to
\bea
\chi + {^{40}{\rm Ar}} \to \,{^{41}{\rm Ar}}^* \to \,{^{41}{\rm Ar}} + \gamma_c \ ,
\eea
where $\gamma_c$ denotes cascades of photons. 
The neutron separation energy for the $^{41}{\rm Ar}$ nucleus is  $S(n)= 6.099\ {\rm MeV}$ \cite{Wang_2017}, which implies the total energy per cascade 
\bea
E_c =6.099\ {\rm MeV} - \Delta m \ , 
\eea
where $\Delta m = m_n - m_\chi$. It follows from Eq.\,(\ref{one2}) that
\bea
4.527 \ {\rm MeV} < E_c < 5.317 \ {\rm MeV} \ .
\eea

Provided the decay pattern of the $\chi$ capture state is similar to that of the neutron capture state, the energies of photons in the cascades can be determined from the known data for neutron capture by $^{40}{\rm Ar}$ \cite{LYCKLAMA196733,Hardell_1970,STE}. The only difference is that  the energy of the first  photon in each cascade  from dark matter capture will be smaller by $\Delta m $ than the energy of the corresponding photon in a neutron capture process.

\begin{table}[t!]\label{tabel1}
\begin{center}
\begingroup
\setlength{\tabcolsep}{6pt} 
\renewcommand{\arraystretch}{1.5} 
\begin{tabular}{  |c | l | c |} 
\hline
Br  & \ \ \ \ \ \ \ \ \ \  Photon energies  [MeV] \ \ & {$\sigma_{n^*}^{\rm NR}/\sigma_{n}^{\rm NR}$} \\ [2pt]
\hline\hline
 $31\%$  &  {$4.745 \!-\! \Delta m$}\,, \     {$1.187$}\,, \   {$0.167$}   & {0.64 -- 0.82} \\ [2pt]
\hline
$9\%$ & {$5.583 \!-\! \Delta m$}\,, \   {$0.516$}  &  {0.68 -- 0.84} \\ [2pt]
\hline
$6\%$ &    {$4.745 \!-\! \Delta m$}\,, \ {$0.838$}\,, \     {$0.516$} &  {0.64 -- 0.82}  \\ [2pt]
\hline
$6\%$ &   {$2.772 \!-\! \Delta m$}\,, \   {$2.811$}\,, \   {$0.516$}    &   {0.42 -- 0.70}   \\ [2pt]
\hline
$4\%$ &  {$3.701 \!-\! \Delta m$}\,, \   {$1.044$}\,, \  {$1.187$}\,, \  {$0.167$}   &  {0.55 -- 0.77}  \\
\hline
\end{tabular}
\endgroup
\end{center}
\vspace{-1mm}
\caption{The photon energies for the dominant cascades from dark matter capture by $^{40}{\rm Ar}$, along with  their branching fractions (Br)  and the range of possible values of the ratio $\sigma_{n^*}^{\rm NR}(^{40}{\rm Ar})/\sigma_{n}^{\rm NR}(^{40}{\rm Ar})$ for each cascade separately; $\Delta m = m_n-m_\chi$.} \label{tabel1}
\end{table}

The energies of photons in the five dominant cascades from dark matter capture by $^{40}{\rm Ar}$ are summarized  in Table \ref{tabel1} along with their branching fractions.
Cascades that are not listed  have branching fractions less than $2\%$.
Only two $^{41}{\rm Ar}$ excited states have known lifetimes: $0.167 \ {\rm MeV}$ $(\tau = 0.32 \ {\rm ns})$ and $0.516 \ {\rm MeV}$ $(\tau = 0.26 \ {\rm ns})$; lifetimes of the other excited states of $^{41}{\rm Ar}$ are likely also shorter than $1 \ {\rm ns}$. Table  \ref{tabel1} provides also the suppression factors $\sigma_{n^*}^{\rm NR}(^{40}{\rm Ar})/\sigma_{n}^{\rm NR}(^{40}{\rm Ar})$ due to the lower energy of the first photon in each cascade compared to the neutron capture case. 
To compute those factors, an analogous formula to the one in Eq.\,({\ref{nstar}}) was applied to each channel of dark matter capture by argon, with the adopted parameter values for argon $R\approx 3.43 \ {\rm fm} $ \cite{ANGELI201369} and $a_s \approx 1.83 \ {\rm fm}$ \cite{neutrondata}.

 Although there are no resonances for neutron capture by argon at energies $\sim 530 \ {\rm eV}$, a resonance behavior may arise in the case of dark matter capture. However, as discussed earlier, due to the expected narrowness  and  low density of nuclear states in argon, as well as the broad velocity distribution of dark matter, the resonant contribution will not change the cross section  significantly.  
 
 Given the comparable sizes of the suppression factors $\sigma_{n^*}^{\rm NR}/\sigma_{n}^{\rm NR}$ for each cascade, in order to estimate the rate of dark matter capture  at DUNE we assume $\sigma_{n^*}^{\rm NR}(^{40}{\rm Ar})\approx 0.7\,\sigma_{n}^{\rm NR}(^{40}{\rm Ar})$ for the total cross section. This gives
\bea
R_{\rm DUNE}  \ &\approx& \ 0.02  \left[\frac{\varepsilon}{m_n-m_\chi}\times 10^{10}\right]^2  \ \ {\rm  events}/{\rm day} \ . \ \ \ \ \ \ 
\eea
Based on the constraints shown in Figure \ref{fig1}, the largest allowed value of $\varepsilon/(m_n-m_\chi)$ is $\sim5\times 10^{-10}$, which yields
a dark matter capture rate at DUNE of $\sim 0.5$  events/day.

Such an event rate with its unique signature is very nontrivial. However, whether it can be distinguished from the background depends on the progress of several ongoing research and technological development efforts at DUNE and the final experimental setup. In particular, the optimal photon detection strategy is still under investigation.

An important background comes from the neutron capture process, where  the neutrons originate from the environment. This can be significantly reduced by implementing a hermetic passive water shield as suggested in Ref.\,\cite{Capozzi:2018dat}. 
An excellent  resolution of the photon energy can also be used to  reduce this background, since the first photon in each cascade has a significantly lower energy than the corresponding photon from a neutron capture process. 

A possible complication arises from the fact that it might not be easy to distinguish an electron from a photon at DUNE, especially in the low-energy regime. Thus, a $\beta$ decay of a radioactive nucleus can be an important contribution to the background \cite{Capozzi:2018dat}. A good particle identification efficiency is required to reduce such a background.  

A detailed experimental analysis is required before making definitive claims regarding the detectability prospects of dark matter capture at DUNE. 
If the signal is indeed observable, it would be interesting to investigate its annual modulation due to the possible  nontrivial resonant structure.

\subsection*{PandaX,\, XENON1T,\, LUX}

Several experiments with detectors based on  xenon time projection chamber technology are searching for dark matter  nuclear recoils. They are using natural xenon, for which the most abundant isotopes are:  $^{129}{\rm Xe}$ $(26.4\%)$, $^{131}{\rm Xe}$ $(21.2\%)$ and $^{132}{\rm Xe}$ $(26.9\%)$. In their final runs PandaX \cite{Cao:2014jsa} and LUX \cite{Akerib:2012ys}  had fiducial volumes of $\sim 300 \ {\rm kg}$, whereas XENON1T \cite{Aprile:2015uzo} had $\sim 2 \ {\rm tons}$ of xenon. 

Let us consider the isotope $^{131}{\rm Xe}$. 
In the kinetic energy region $E_k \sim 530 \ {\rm eV}$, the neutron capture cross section, with a nonresonant contribution $\sigma_n^{\rm NR}(^{131}{{\rm Xe}})\sim 1 \ {\rm b}$, exhibits a resonant behavior. According to theoretical predictions, it can reach up to $\sim700 \ {\rm b}$ at the resonance peak, but the width of the resonances is small \cite{etde_519804}.

The dark matter capture by $^{131}{\rm Xe}$ would result in  cascades of photons  through the process
\bea
\chi + {^{131}{\rm Xe}} \to \,{^{132}{\rm Xe}}^* \to \,{^{132}{\rm Xe}} + \gamma_c \ .
\eea
The neutron separation energy for the final state $^{132}{\rm Xe}$ is $S(n)=8.937 \ {\rm MeV}$, thus the total energy of the cascade would be $E_c = 8.937 \ {\rm MeV} - \Delta m$, i.e.,
\bea\label{ranE}
7.365 \ {\rm MeV} < E_c < 8.155 \ {\rm MeV}\ . 
\eea
In the case of xenon, the structure of the photon cascades is very complex. 
To arrive at a rough estimate for the dark matter capture rate, we approximate $\sigma_{n^*}^{\rm NR}(^{131}{\rm Xe})\approx \sigma_{n}^{\rm NR}(^{131}{\rm Xe}) \sim 1 \ {\rm b}$. The nonresonant contribution leads to 
\bea
R_{\rm Xe}  \ &\sim& \ 10^{-5}  \left[\frac{\varepsilon}{m_n-m_\chi}\times 10^{10}\right]^2  \ \ \frac{\rm  events}{\rm ton\times day} \ , \ \ \ \ \ \ \ 
\eea
which, for the largest experimentally allowed value of $\varepsilon/(m_n-m_\chi) \sim 5\times 10^{-10}$, gives only $\sim 10^{-4}$  events/ton/day.

At PandaX-II  a detailed evaluation of the neutron background has been performed \cite{Wang:2019opt}. 
It was determined for two low-background runs:
Run $9$ ($26$-ton-day) and Run $10$ ($28$-ton-day)  \cite{Wang:2019opt}.   
The estimated neutron capture rate was $\sim 0.7$ events/ton/day in Run $9$ and $\sim 0.4$ events/ton/day in Run $10$. Those signals came from neutron capture 
mostly on $^{129}{\rm Xe}$ and $^{131}{\rm Xe}$, since those two isotopes have the largest cross sections for thermal {$(\sim 25 \ \rm meV)$} neutron capture: $\sim 20 \ {\rm b}$ and $100 \ {\rm b}$, respectively. 
The corresponding 
total energies of the resulting photon cascades (the current PandaX detector is  capable of measuring only the total energy deposition and cannot distinguish each photon), equal to the neutron separation energies in $^{130}{\rm Xe}$ and $^{132}{\rm Xe}$, are $9.256 \ {\rm MeV}$ and $8.937 \ {\rm MeV}$. The capture by other, less abundant xenon isotopes produce cascades with energies $<6.5 \ {\rm MeV}$.

The current energy resolution at PandaX is about $5\!-\!10\%$. The nonresonant dark matter capture by xenon gives a signal rate which is a few orders of magnitude lower than  the neutron capture background, making it difficult to tell them apart. However, the energy resolution is planned to be improved in these experiments in the $\sim 10 \ {\rm MeV}$ energy regime. In addition, as estimated in Sec.\,\ref{sec3}, the resonant contribution to the capture cross section can be comparable or even larger (up to a factor of $\sim 10$) than that of the nonresonant channel, making it easier to distinguish the signal from the background. Since the resonance channel has a sensitive velocity dependence, this may also lead to a significant annual modulation of the signal.

Combining a better energy resolution, an enhanced capture cross section from resonance effects with an annual modulation feature, as well as a much larger amount of data to be collected, future experiments, such as PandaX-4T, XENONnT and LUX-Zeplin, may provide promising probes of the dark matter capture process.

\section{Discussion}

We have proposed a new method to search for a particular kind of dark matter particle  based on its capture by atomic nuclei. The signature of such a process is a single photon or a cascade of photons resulting from the de-excitation of the capture state. The energy of this single photon or  the total energy of the cascade (and of the first photon in each cascade) distinguishes the signal from the neutron capture background.

The ideal places to look for dark matter capture are large volume neutrino experiments and dark matter direct detection experiments, especially those with scintillation detectors and time projection chamber  technology. Whether the signal\break can stand out from the background, mainly from natural\break neutron capture, depends on the detailed design of each experiment. The ongoing research and development  at DUNE, and future improvements at various xenon experiments, show great promise in searching for dark matter capture. A detailed analysis of the event reconstruction efficiency, object identification capability and detector resolution are still needed to reach a concrete conclusion.

In some  liquid scintillator experiments, like KamLAND, which measures the antineutrino flux from nuclear reactors, a delayed coincidence cut is imposed to reduce the background  \cite{Eguchi:2002dm}. This cut removes the potential signal from dark matter capture.
 It would be interesting to perform an analysis of the photon background with this cut abandoned.

It might be worth exploring the possibility of using other detector materials with large neutron capture cross sections, e.g., iodine and samarium \cite{etde_519804}.  Out of those,  iodine is already\break widely used as a scintillator in the form of sodium iodine or cesium iodine. 
Further examples of nuclei with large neutron\break capture cross sections are terbium and gadolinium, which could also be used to construct future detectors with increased sensitivity to dark matter capture. Given the expected nontrivial resonance structure of the nuclear states, an annual\break modulation of the signal may be observed.
A more economical solution would be to dope existing detectors with such substances. 
This has  been considered in a different context for the case of  gadolinium at Super-Kamiokande \cite{Beacom:2003nk,Renshaw:2012np}.

Finally, we would like to emphasize that the dark matter capture explored
in this letter offers an alternative way to look for the
neutron dark decay channel $n\to \chi\gamma$, with a similar or better sensitivity than direct searches.\vspace{2mm}

\subsection*{Acknowledgments}

We would like to thank Jianglai Liu, David McKeen, Maxim Pospelov, Yun-Tse Tsai, Donglian Xu and Bei Zhou for helpful discussions.
B.F. and Y.Z. are supported by the U.S. Department of Energy under Award
No.~${\rm DE}$-${\rm SC0009959}$. B.G. is supported by the U.S. Department of Energy Grant No.~${\rm DE}$-${\rm SC0009919}$.

\bibliography{DM}

\end{document}